\documentclass[sigplan,screen,natbib]{acmart}
\AtBeginDocument{%
  }

\usepackage{tikz}
\usepackage{booktabs}
\usepackage{tabularx}     
\usepackage{makecell}     
\usepackage{threeparttable}
\usepackage{amsmath}

\renewcommand\footnotetextcopyrightpermission[1]{} 
\DeclareMathOperator*{\argmin}{arg\,min}

\selectfont

\newcommand{\circnum}[1]{%
  \tikz[baseline=(n.base)]\node[draw,circle,inner sep=1pt,minimum size=1em](n){\footnotesize #1};%
}

\AtBeginDocument{
  \setlength{\abovedisplayskip}{1pt}
  \setlength{\belowdisplayskip}{1pt}
  \setlength{\abovedisplayshortskip}{2pt}
  \setlength{\belowdisplayshortskip}{5pt}
}
\usepackage{booktabs}
\begin{document}

\title[ROSA: Robust and Energy-Efficient Microring-Based Optical Neural Networks]{ROSA: \underline{R}obust and Energy-Efficient Microring-Based Optical Neural Networks via \underline{O}ptical \underline{S}hift-and-\underline{A}dd and Layer-Wise Hybrid Mapping}



\author{
Huifan Zhang$^{1}$, Yun Hu$^{1}$, Caizhi Sheng$^{1}$, Yurui Qu$^{1}$, Pingqiang Zhou$^{1}$
\\[4pt]
$^{1}$ShanghaiTech University, Shanghai, China
}

\renewcommand{\shortauthors}{Zhang et al.}

\begin{abstract}
This work presents ROSA, a microring-based optical neural network architecture that improves robustness and energy efficiency using an optical shift-and-add (OSA) module and a layer-wise hybrid mapping strategy. It introduces a noise-aware voltage-to-weight model considering DAC and thermal variations, and a workload-aware framework to co-optimize MRR array size and layer-wise dataflow. Optimized arrays reduce the aggregated relative energy-delay product (EDP) by $64\%$ and $26\%$ compared with DEAP-CNNs and a general compact array, respectively. OSA further contributes $29\%$ EDP reduction. The proposed hybrid mapping strategy improves CIFAR-10 accuracy by $8.3\%$ over weight-stationary mapping while achieving an average $54.7\%$ lower EDP than DEAP-CNNs.

\end{abstract}




\maketitle


\section{Introduction}

The emergence of large language models (LLMs) and deep neural networks (DNNs) has driven an increasing demand for high-throughput and energy-efficient inference accelerators. However, traditional electronic architectures struggle to meet these demands because of the von Neumann bottleneck and limited parallelism. As a result, alternative computing paradigms have been actively explored to overcome these limitations. Optical neural networks (ONNs) have gained significant attention due to their potential for ultra-high bandwidth and low latency computation \cite{hua2025integrated, ahmed2025universal, zhou2025toward, liu2023fiona, zhu2022space, yin2025toward}. Among them, integrated microring-resonator (MRR)-based ONNs~\cite{tait2017neuromorphic, tait2014broadcast} have emerged as a promising candidate. Their compact footprint, low latency enabled by the inherent speed of light, and massive parallelism enabled by dense wavelength-division multiplexing (WDM) make them highly attractive for accelerating modern deep neural networks. 

Several recent works have explored MRR-based ONN architectures \cite{bangari2019digital, liu2019holylight, sunny2021crosslight, gu2022squeezelight}. DEAP-CNNs~\cite{bangari2019digital} proposes an electro-optic architecture that utilizes MRRs to perform multiply-and-accumulate (MAC) operations, demonstrating significant speedup and energy savings compared with GPUs. However, this work modulates both inputs and weights in analog format, and thus suffers from the slow thermal-optic (TO) tuning bottleneck and noise-induced accuracy degradation. HolyLight~\cite{liu2019holylight} attempts to bypass this bottleneck by using fast electro-optic (EO) tuned microdisk as computing units. CrossLight~\cite{sunny2021crosslight} utilizes static TO tuning for wide weight range and a dynamic EO tuning over a smaller range. SqueezeLight~\cite{gu2022squeezelight} further proposes a device-level, innovative MRR design with multiple operands to improve throughput. However, all these works face a difficult dilemma: either rely on slow TO tuning for analog weight updates or sacrifice high throughput using digital EO encoding. 

\begin{figure}[tbp]
\centerline{\includegraphics[scale=0.6]{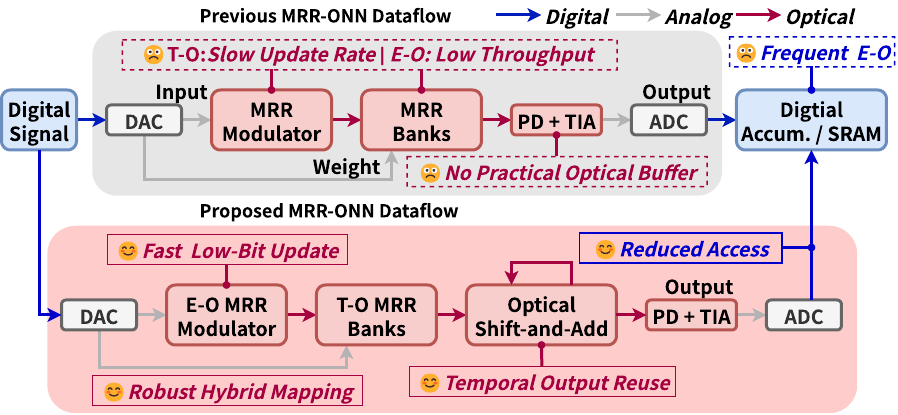}}
\caption{Previous and proposed MRR-ONN dataflow.}
\label{fig:simple_flow}
\end{figure}

Additionally, the nature of photons makes it difficult to store and buffer the outputs. In prior MRR-ONN architectures, there is a lack of output reuse mechanisms, as illustrated in Fig.~\ref{fig:simple_flow}, a conventional MRR-ONN repeatedly converts partial sums between the optical and electrical domains, and stores them only in electronic buffers. For EO-tuned MRRs, the overhead of optical-to-analog conversion (OAC) and analog-to-digital conversion (ADC) is even worse. 

In this paper, we propose ROSA, a robust and energy-efficient MRR-ONN architecture that addresses the aforementioned challenges through three key methods: (1) reducing expensive digital-analog-optical conversions by performing as much accumulation as possible in the optical domain with an optical shift-and-add (OSA) module, (2) alleviating the slow TO tuning bottleneck while utilizing its high throughput by representing analog weights with a digital-analog computing mode, and (3) improving robustness against thermal noise to enable high-bit quantization for analog MRR values with a layer-wise hybrid mapping strategy. Our main contributions are as follows:
\begin{itemize}
  \item We propose a MRR-ONN architecture with the OSA module that enables optical signal accumulation to reduce the frequency of costly OAC/ADC conversions. This design significantly reduces the energy delay product (EDP) by $29\%$ compared to the non-OSA structure.
  \item We optimize the dimensions of the MRR array for different DNN workloads, reducing the aggregated relative EDP by $63.7\%$ and $26\%$ compared to the DEAP-CNNs \cite{bangari2019digital} setting and the compact general arrays $4\times 4$ \cite{tait2017neuromorphic}, respectively.
  \item We develop a layer-wise hybrid-mapping strategy that optimizes robustness and energy efficiency. Tested on representative 8-bit quantized CNNs, it has only 3.3\% accuracy loss on CIFAR-10 compared to the ideal full-precision model, while reducing EDP by $54.7\%$ averagely compared to DEAP-CNNs. 
\end{itemize}

\begin{figure*}[tbp]
\centerline{\includegraphics[width=\textwidth]{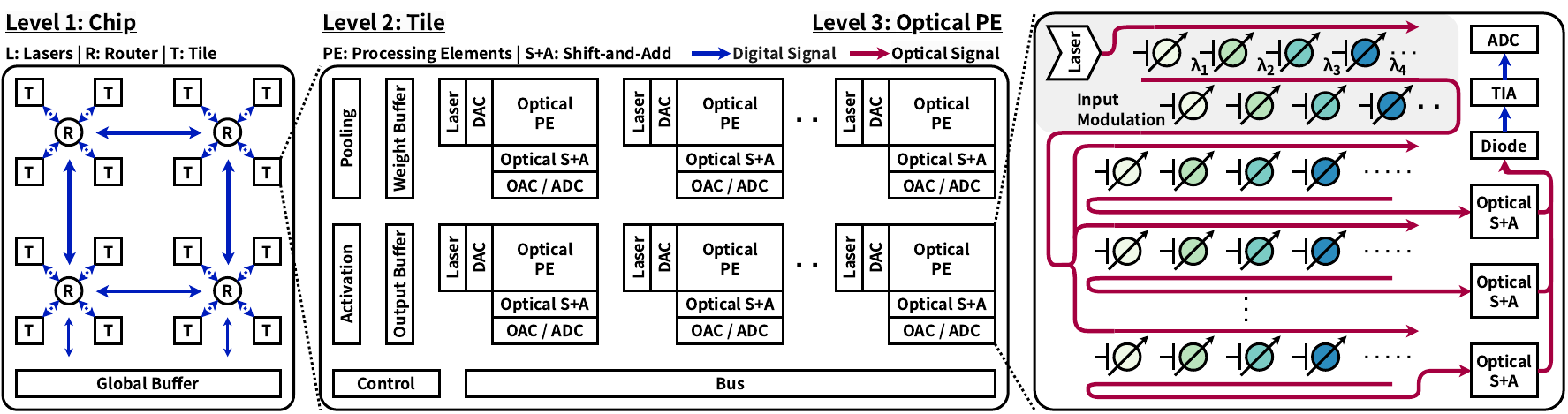}}
\caption{Architecture of the proposed MRR-based ONN accelerator with optical shift-and-add module.}
\label{fig:Architecture}
\end{figure*}

\section{Preliminary and Motivation}
In this work, we focus on MRR-based ONNs that encode signals in the amplitude of optical waves.
Using wavelength-division multiplexing (WDM), multiple signals at different wavelengths can be processed in parallel. The weights are mapped to the transmission response of MRRs. The multiplication between input and weight is performed as the input amplitude attenuated by weight MRR transmission. After photo detection, the optical signals are converted to electrical domain and summed up to generate the output. Assume the input signal power is modulated to \( x_k \) on each of the \( N_\lambda \) wavelength channels \( \{\lambda_k\}_{k=1}^{N_\lambda} \). Each MRR implements a weight \( w_k \), represented by its power transmission. The differential drop-trough transmission is used to ensure a full range mapping. The output is a value given by \( y = \sum_{k=1}^{N_\lambda} w_k x_k \). This process realizes the fundamental multiply-and-accumulate (MAC) operation in neural networks supporting both convolutional layer and general matrix multiplication (GEMM).


Based on this principle, computation can operate in either a \textbf{digital} or an \textbf{analog} mode. In the digital mode, input signals are encoded in binary format \cite{zhu2024lightening,liu2019holylight}, whereas in the analog mode, inputs are represented as continuous values \cite{tait2017neuromorphic,sunny2021crosslight,gu2022squeezelight,bangari2019digital}.  The analog mode typically achieves a higher total operations per second than the digital mode, since multiple bits can be processed per cycle, but it is also more susceptible to noise and device non-idealities under high resolution. Fine-grained dynamic calibration of the transmission relies on thermal tuning, where the heating and cooling process usually takes 5-10~$\mu$s~\cite{liu2022thermo}, limiting the weight-update rate to roughly 500~kHz. By comparison, the EO tuning frequency used in digital mode can reach 48.6~GHz \cite{yuan20245}. 

To alleviate this bottleneck while preserving the high OPS of analog computing, we observe that a mixed digital-analog computing paradigm that keeps weights in analog format while encoding inputs digitally is simple yet effective. This approach leverages high-speed EO modulation for input encoding, bypassing the slow TO tuning. Table.~\ref{tab:comparison} compares different modes of an $R\times C$ MRR array which quantize input and output into $N_i$ and $N_w$ bits, respectively. Compared to the fully analog mode, it reduces the update time from $\mu s$ to $ps$ scale, thus significantly improving the OPS. Compared to the digital mode, it retains the high throughput advantage by keeping weights in analog format.

\begin{table}[thbp]
\small                                     
\centering
\caption{Comparison of our proposed MRR-ONN with prior works.}
\label{tab:comparison}
\begin{tabularx}{\linewidth}{lccc}   
\toprule
Mapping Mode & Analog & Digital & Mixed \\
\midrule
Example & DEAP-CNNs \cite{bangari2019digital} & HolyLight \cite{liu2019holylight}  & Proposed \\
\midrule
Throughput & $R C N_i  N_w$ & $R C$ & $R C N_w$ \\
Update Time & $t_{TO}$ (5-10~$\mu$s) & $t_{EO}$ (20-40~ps)  & $t_{EO}$ \\
OPS & $R C N_i  N_w / t_{TO}$ & $R C / t_{EO} $ & $R C N_w / t_{EO}$ \\
Robustness & Low & High & High \\
OADC Energy & Low & High & Low \\
\bottomrule
\end{tabularx}
\end{table}

\section{ROSA: Architecture, Modeling, and Optimization}

Figure~\ref{fig:Architecture} gives a high-level view of the proposed MRR-ONN accelerator. It adopts a hierarchical architecture with three levels: \circnum{1} \textbf{chip}, \circnum{2} \textbf{tile}, and \circnum{3} \textbf{optical processing elements (OPEs)}. At the \textit{chip level}, a router-connected mesh of tiles exchanges input, weight, and output data with off-chip DRAM, and a global buffer caches input and output tensors. At the \textit{tile level}, OPEs execute matrix-vector multiplications together with activation and pooling for conventional CNNs. On-chip weight buffers store the bias voltages of the weight MRRs, and output buffers accumulate partial sums. Each tile integrates a laser module to illuminate the OPE inputs and a DAC to drive the MRR modulation voltages. On the output side, each OPE is attached to an optical shift-and-add module followed by OAC and ADC stages. At the \textit{OPE level}, illustrated in the grey shaded area, paired parallel MRRs modulate \(N_{\lambda}\) wavelength groups of the input laser to realize input encoding. The modulated signal is broadcast to \(N_{r}\) rows of MRR arrays for input reuse; each row performs the MAC operation in~(\ref{eq:MAC}). Per-slot products are accumulated optically via OSA; the summed photocurrent is amplified by a TIA and digitized by an ADC. 

This hierarchical picture provides the system context for the subsequent sections: Section 3.1 details the OSA module, Section 3.2 explains the computational mapping on this architecture, and Section 3.3 to 3.5 develop noise and energy models for design-space exploration. 

\subsection{Optical Shift-and-Add Module}

\begin{figure}[htbp]
\centerline{\includegraphics[scale=0.6]{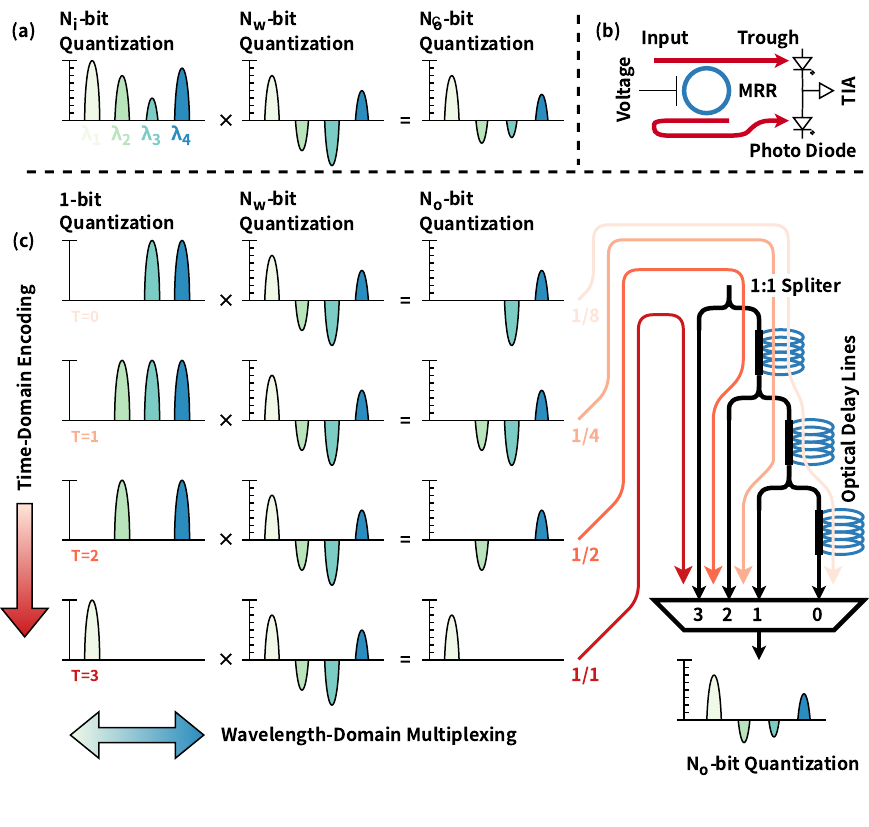}}
\vspace{-0.3in}
\caption{MAC computing paradigms using integrated MRRs and optical shift-and-add module. 
\textbf{(a)} Conventional broadcast-and-weight mode. 
\textbf{(b)} The basic MRR MAC unit.
\textbf{(c)} Proposed optical shift-and-add module.}
\label{fig:coding}
\end{figure}

Under the digital-analog paradigm, storing and reusing output partial sums remains challenging. Nevertheless, they can be temporally buffered at very low propagation loss using optical delay lines (ODLs). In this paper, as shown in Fig.~\ref{fig:coding}(c), we propose a pure optical module which enables output shift-and-add operations and thus reduces the energy consumption of optical-to-analog conversion (OAC) and analog-to-digital conversion (DAC). The \textit{shift} operation is implemented by light splitters and ODLs, while the \textit{add} operation is performed by photo-detection and trans-impedance amplification (TIA).

A \textbf{light splitter} divides an optical signal into specified ratios. A 1:1 splitter, for instance, performs a divide-by-two operation on the analog optical power. Sugeet Sunder et al. \cite{sunder2025scalable} proposed an analog-digital optical computing module that converts the optical power into an N-bit digital representation through a series of such splitters. \textbf{ODLs} adjust the temporal sequence of optical signal streams, aligning them accurately at the output port. Among optical integrated implementations, the state-of-the-art SCISSOR delay line \cite{morton2011fast} achieves a tunable delay of up to 345 ps, corresponding to a \textit{minimum} input signal frequency of 2.9 GHz, which is lower than the EO tuning frequency. Therefore, an operating frequency of 5-10 GHz is feasible for the proposed OSA module. ODLs are not a free lunch: if each slot has delay variation, the final accumulated output becomes inaccurate. To mitigate this, we can employ active calibration with integrated phase modulators \cite{marandi2014network}. 

This OSA module naturally supports digital-analog MAC operations. Assume a \textit{normalized} input \( x_k \in (-1,1) \) is quantized into \( N_T \) bits using balanced ternary symbols \( b_{k,t} \in \{0,-1,1\} \), where \( b_{k,0} \) and \( b_{k,T} \) denote the least and most significant bits, respectively. At each time step \( t \), the instantaneous output \( y_{k,t} = w_k b_{k,t} \) is proportionally scaled by \( 2^{-(N_T - t)} \) during the final detect-and-sum cycle. Consequently, the accumulated optical output
\begin{align}
    y &= \sum_{k=1}^{N_\lambda} \sum_{t=0}^{N_T} 2^{t - N_T} w_k b_{k,t} \\
      &= \sum_{k=1}^{N_\lambda} w_k \sum_{t=0}^{N_T} 2^{t - N_T} b_{k,t} = \sum_{k=1}^{N_\lambda} w_k x_k,
      \label{eq:MAC}
\end{align}
is mathematically equivalent to that produced by the conventional broadcast-and-weight method. 
\textit{It is worth noting that} the OSA module supports not only ternary coding, but also pulse amplitude modulation (PAM) with higher bitwidths, which can further improve the throughput.

\subsection{Mapping Strategies}

\begin{figure}[htbp]
\centerline{\includegraphics[scale=0.6]{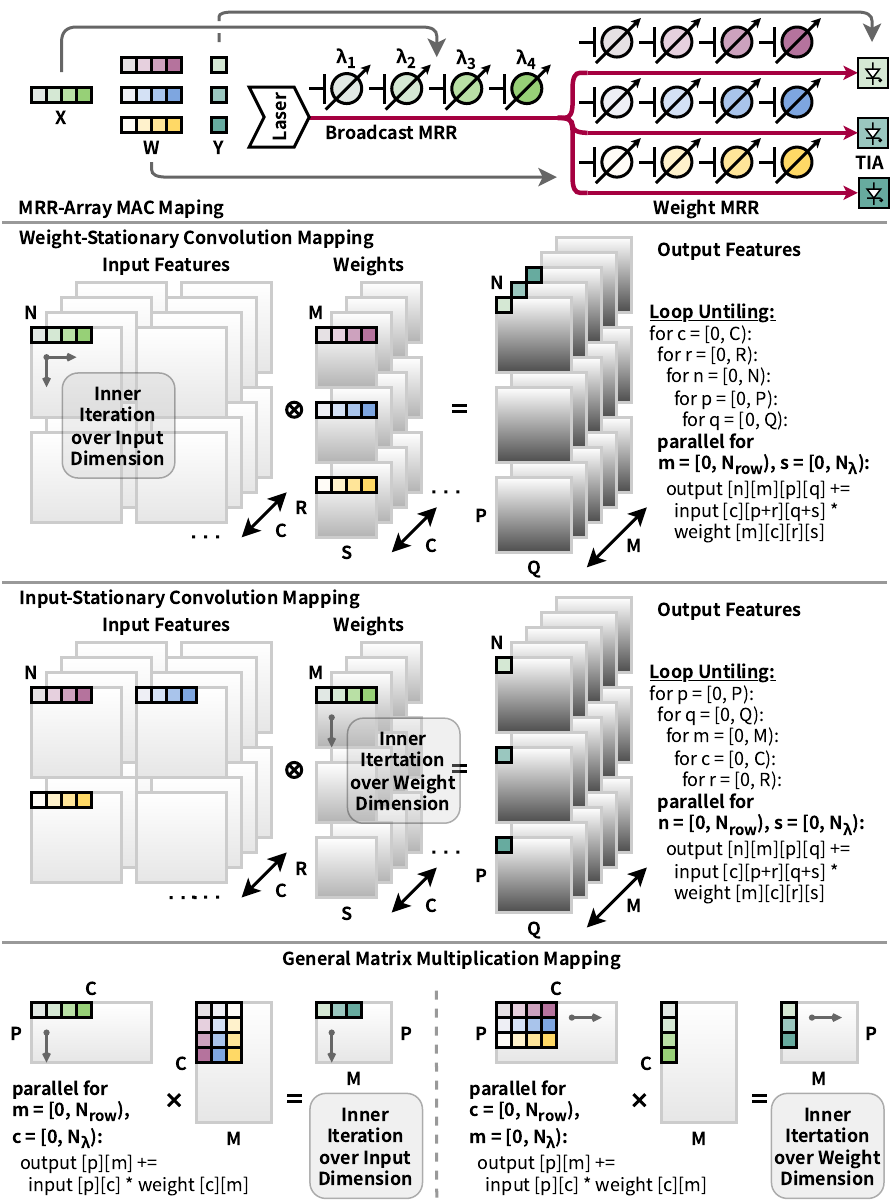}}
\caption{Different mapping strategies for MRR-ONN.}
\label{fig:mapping}
\end{figure}

\noindent\textbf{Convolution mapping.} Previous works on MRR-ONN usually use \textit{weight-stationary} scheme for full convolution layers: the bias voltages on the weight MRRs are held constant while the inner loops iterate over input features (Fig.~\ref{fig:mapping} top). 
In contrast, an \textit{input-stationary} scheme where input features are programmed onto the weight MRRs and the kernel weights are encoded via the broadcast MRRs has been largely underexplored (Fig.~\ref{fig:mapping}, middle). Although it typically incurs higher energy due to more frequent weight modulation and broadcast, the sensitivity to noise of input features is lower than that of weights in some layers. Thus, input-stationary mapping can improve robustness, as we will analyze in Section 3.5. 

\noindent\textbf{GEMM mapping.} With the growing demands of Transformer workloads, large-scale general matrix multiplication (GEMM) for query-key/value projections must be accelerated. Our architecture fully utilizes the optical MAC unit to map GEMM (Fig.~\ref{fig:mapping}, bottom): multipliers are parallelized across $N_{\mathrm{row}}$ OPE rows, while the inner loop accumulates the vector multiplication result across $N_{\lambda}$ wavelength groups.

\subsection{Noise-Aware Behavioral Modeling of MRR Weight Realization}
\begin{figure}[hhtbp]
\centerline{\includegraphics[width=\linewidth]{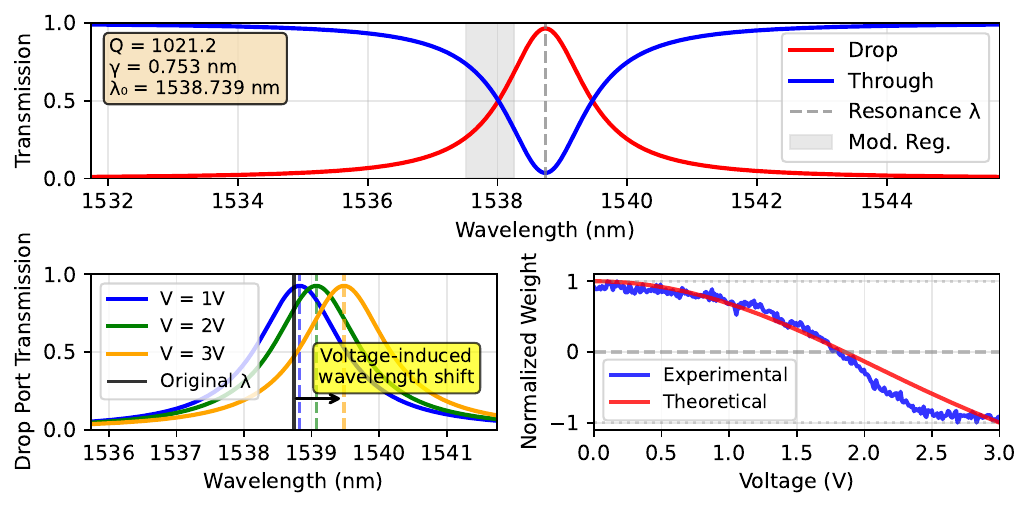}}
\caption{\textbf{(a)} MRR transmission characteristics around $\lambda_0$ = 1538.739 nm. 
\textbf{(b)} Voltage tuning from 1V to 3V, max wavelength shift: 0.740 nm.
\textbf{(c)} Experimental vs theoretical correlation.}
\label{fig:MRR-ONN}
\end{figure}

Since the output and weight of our proposed MRR-ONN are both in analog format, the robustness against noise largely determines the available quantization levels and thus the overall accuracy. As the cornerstone of noise-aware behavioral model, it is crucial to derive the mapping from applied voltage to effective weight values.
First, for TO-tuned MRRs, the applied voltage \( V \) generates a temperature change \( \Delta T \) in the heater, which induces a resonance wavelength shift \( \Delta \lambda \):
\begin{equation}
  \Delta T(V) = \frac{V^2}{R_h} R_{\mathrm{th}}, \quad \Delta\lambda(\Delta T) = \lambda_0 \frac{\beta\Delta T}{n_0+\beta\Delta T}
\label{eq:tow}
\end{equation}
where \( R_h \) is the heater resistance, \( R_{\mathrm{th}} \) is the thermal resistance, \( \lambda_0 \) is the nominal resonance wavelength, \( n_0 \) is the effective refractive index, and \( \beta \) is the thermo-optic coefficient. Next, the drop-port transmission at a fixed probe wavelength \( \lambda_{\mathrm{ref}}\) is given by:
\begin{equation}
  T_{\mathrm{drop}}(\lambda, \lambda_{\mathrm{ref}}) = \frac{\gamma^{2}}{\bigl(\lambda-\lambda_{\mathrm{ref}}\bigr)^{2} + \gamma^{2}}
\end{equation}
where \( \gamma \) is the half-width at half-maximum (HWHM) of the resonance. The weight is normalized to the differential transmission between the drop and through ports:
\begin{equation}
  T_{\mathrm{diff}} = T_{\mathrm{drop}} - T_{\mathrm{thru}} =  T_{\mathrm{drop}} - (1 -  T_{\mathrm{drop}})
  \label{eq:lz}
\end{equation}
\begin{equation}
  T_{\mathrm{hi}} = T_{\mathrm{diff}}(V_{\min}),\ T_{\mathrm{lo}} = T_{\mathrm{diff}}(V_{\max})
\label{eq:bounds}
\end{equation}
\begin{equation}
  w(V) = Q_{\min} + Q_{\mathrm{rng}} \frac{T_{\mathrm{diff}}(V)-T_{\mathrm{lo}}}{T_{\mathrm{hi}}-T_{\mathrm{lo}}}
  \label{eq:wfromT}
\end{equation}
where $Q_{\min}$ and $Q_{\max}$ are the lower and upper bounds of the weight range, respectively, and $Q_{\mathrm{rng}} = Q_{\max} - Q_{\min}$. The MRR transmission characteristics with no voltage bias are shown in Fig.~\ref{fig:MRR-ONN}(a). By applying a voltage from 1V to 3V, the resonance wavelength shifts by up to 0.740 nm, as shown in Fig.~\ref{fig:MRR-ONN}(b). The experimental and theoretical correlation of voltage-to-weight mapping is illustrated in Fig.~\ref{fig:MRR-ONN}(c). Symbols and parameter values are summarized in Table~\ref{tab:mrr_params}. 






\begin{table}[thbp]
\small                                     
\centering
\caption{Symbols and parameter values for the microring and thermo-optic model.}
\label{tab:mrr_params}
\begin{tabularx}{\linewidth}{clcc}   
\toprule
\textbf{Symbol} & \textbf{Description} & \textbf{Value} & \textbf{Unit} \\
\midrule
$\lambda_0$ & Nominal resonance wavelength & 1538.74 & nm \\
$\lambda_{\mathrm{ref}}$ & Probe (reference) wavelength & 1538.26 & nm \\
$a$ & Round-trip attenuation factor & 0.925 & — \\
$n_0$ & Effective refractive index & 2.4 & — \\
$\gamma$ & Half-width at half-maximum & 0.7534 & nm \\
\midrule
$R_h$ & Heater resistance & 50 & $\Omega$ \\
$R_{\mathrm{th}}$ & Thermal resistance & 2 & $\mathrm{K}/\mathrm{mW}$ \\
$\beta$ & Thermo-optic coefficient & $1.86\times10^{-4}$ & $\mathrm{K}^{-1}$ \\
\bottomrule
  \end{tabularx}
  \end{table}

Based on the above modeling, any perturbation along the chain \(V \!\rightarrow\! \Delta T \!\rightarrow\! \Delta\lambda \!\rightarrow\! T(\lambda_{\mathrm{ref}}) \!\rightarrow\! w\) will change the final mapped values. In this paper, we \textit{currently} consider two noise sources: (i) \emph{DAC-induced} imperfections that shift the applied voltage \(V\), and (ii) \emph{thermally induced} crosstalk that distorts the effective refractive index. In what follows, both are modeled as zero-mean Gaussian noise terms added at their respective stages, i.e., \(\varepsilon_{\mathrm{DAC}}\sim\mathcal{N}(0,\sigma_{\mathrm{DAC}}^{2})\) and \(\varepsilon_{\mathrm{th}}\sim\mathcal{N}(0,\sigma_{\mathrm{th}}^{2})\).
\begin{equation}
V' = V + \varepsilon_{\mathrm{DAC}},\  \Delta T'(V') = \tfrac{V'^2}{R_h}R_{\mathrm{th}} + \varepsilon_T.
\end{equation}

The above noise analysis stands for TO-tuned MRRs. For the EO-tuned MRRs which operate via the free-carrier plasma-dispersion effect under reverse-biased PN junctions, their thermal-induced variation can be well calibrated with a single-monitoring scheme \cite{liu2024single}.



\subsection{Energy Analysis of MRR-based ONN}
Prior MRR-ONN studies often treat energy and power as fixed constants, ignoring the effects of data movement and the frequency of MRR reprogramming. In this paper, we develop a detailed energy model that captures both dynamic energy per computation and static leakage power. The primary power consumption in MRRs arises from their tuning and modulation mechanisms \cite{tait2022quantifying}. We model the power required to lock and shift the resonance wavelength as static power, and the power to electrically tune the MRRs as dynamic power.

The thermal tuning efficiency of MRRs $\eta_{\lambda P}$ (in nm/mW) quantifies the wavelength shift per unit of applied heater power. It can be derived from the following equation:
\begin{equation}
\eta_{\lambda P}\;\triangleq\;
\frac{\partial \lambda}{\partial P}
= \lambda_0 \,\frac{\beta}{n_\mathrm{eff}}\, R_\mathrm{th} \approx 0.238\ \mathrm{nm/mW}.
\end{equation}
For TO-tuned MRRs, the resonance shift range is half-width at half-maximum (HWHM) $\gamma$. Therefore, the average tuning power is $0.5 \gamma / \eta_{\lambda P}\,=\,1.58\ \mathrm{mW}$. For electro-tuned MRRs, the average tuning power is directly referenced from prior work~\cite{yuan20245} as $6.3\ \mathrm{fJ}/\mathrm{bit}$. Static and dynamic energy Settings for other components are summarized in Table~\ref{tab:energy_blocks}.


\begin{table}[t]
  \caption{Static/dynamic energy per component.}
  \Description{Each component is listed as a block with separate rows for static power, dynamic energy, and area. Components are separated by midrules.}
  \label{tab:energy_blocks}
  \centering
  \begin{threeparttable}
  \small
  \begin{tabularx}{\columnwidth}{l l X}
    \toprule
    \textbf{Component} & \textbf{Metric} & \textbf{Value} \\
    \midrule
    \textbf{Laser} \cite{descos2013heterogeneously} & Static power & 1.38 mW \\
    \midrule
    \textbf{MRR} & Static T-O power & 1.58 mW \\
                                & Dynamic E-O energy & 6.3 fJ / bit  \cite{yuan20245}\\
    \midrule
    \textbf{DAC} \cite{sedighi20118} & Dynamic energy & 5.2 pJ / bit \\
    \midrule
    \textbf{PD + TIA} \cite{sun2015single} & Dynamic energy & 440 fJ / bit \\
    \midrule
    \textbf{SRAM} \cite{wieckowski2011128kb} & Leakage power & 48.1 pW / bit\\
                      & Dynamic energy & 50 fJ / bit \\
    \bottomrule
  \end{tabularx}
  \begin{tablenotes}\footnotesize
    \item[*] ADC behavior is modeled using a regression-based plug-in approach~\cite{andrulis2024modeling}.

  \end{tablenotes}
  \end{threeparttable}
\end{table}
\subsection{Optimization of MRR-ONN Architecture and Layer-wise Mapping Strategy}

DEAP-CNNs \cite{bangari2019digital} explored two representative array configurations: a wide-kernel setting with $R\!=\!12$, $C\!=\!100$ and a high-channel setting with $R\!=\!113$, $C\!=\!9$. In O-HAS \cite{li2021has}, optimized designs for AlexNet and ZFNet employ baseline OPEs of $R\!=\!32$, $C\!=\!32$ and $R\!=\!128$, $C\!=\!128$, respectively. A large column dimension implies a wide set of WDM channels is used. Excessive WDM channels become impractical because the required channel spacing approaches the resonator linewidth and dispersion tolerance limits. Most fabricated MRR-ONNs adopt compact arrays (e.g., $R\!=\!4$, $C\!=\!4$) \cite{tait2017neuromorphic,zhao2025situ,hu2024quantized}. Beyond physical limits, large OPEs also suffer from poor \emph{utilization} on modern workloads with small kernels (e.g., MobileNet V3). Therefore, we set the dimension constraint of our OPE to be $C\!<=\!8$, and total weighting MRR number on the chip $T\!\times\!R\!\times\!C\!<=\!1024$ for a fair comparison with previous works.

\noindent\textbf{Aggregated Optimization of EDP Across Workloads.} We use EDP as the optimization metric because it jointly captures energy efficiency and latency in a single objective. To avoid over optimizing OPE sizing to any single network, we must choose an aggregated metric across multiple workloads. For network $n$ with layers $\mathcal{L}_n$, let the total product be $\{\mathrm{EDP}_n\}$.  We form a balanced geometric mean score 
\begin{equation}
G=\!\Big(\prod_{n=1}^{N}\mathrm{EDP}_n\Big)^{\!1/N}.
\end{equation}
We also consider the worst case  $W_{\max}=\max_n \mathrm{EDP}_n$ and a balanced term $\lambda\in[0,1]$ to construct the final robust efficiency metric $M=(1-\lambda)\,G+\lambda\,W_{\max}.$ We choose the OPE size that minimizes $M$.
\begin{figure}[htbp]
\centerline{\includegraphics[width=\linewidth]{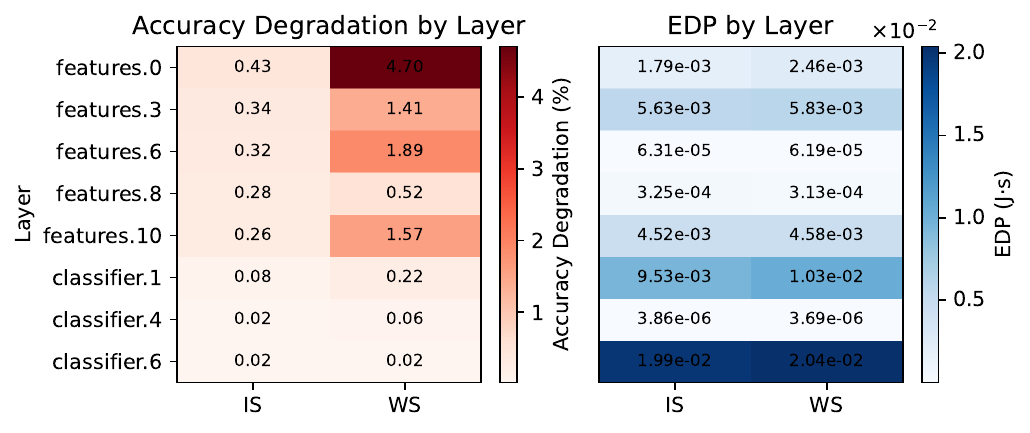}}
\caption{Layer-wise accuracy and EDP comparison between input-stationary and weight-stationary mapping.}
\label{fig:layer-wise}
\end{figure}

\noindent\textbf{Layer-wise Mapping Considering Both Robustness and EDP.} 
As illustrated in Fig.~\ref{fig:layer-wise}, different layers exhibit varying sensitivities to noise and slightly different EDP under IS and WS mappings. To optimally select the mapping for each layer, we propose a balanced metric that considers both. We profile each network layer $l$ under mapping $m\!\in\!\{\mathrm{IS},\mathrm{WS}\}$ on a given dataset to obtain the layer-wise accuracy degradation $d_l(m)$ and EDP $e_l(m)$. An illustrative example using AlexNet on the CIFAR-10 dataset is shown in Fig.~\ref{fig:layer-wise}. We compute a layer-only accuracy weight from the smallest degradation and normalize each candidate by its per-layer best:
\[
d_l^{\mathrm{ref}}=\min\{d_l(\mathrm{IS}),d_l(\mathrm{WS})\},\quad
e_l^{\mathrm{ref}}=\min\{e_l(\mathrm{IS}),e_l(\mathrm{WS})\},
\]
The balanced metric for each mapping is
\[
\argmin_{m} M_l(m) \;=\;
\Big(\tfrac{d_l(m)}{d_l^{\mathrm{ref}}}\Big)^{\alpha_l}\;
\Big(\tfrac{e_l(m)}{e_l^{\mathrm{ref}}}\Big)^{1-\alpha_l},
\qquad m\in\{\mathrm{IS},\mathrm{WS}\}.
\]
where the accuracy weight $\alpha_l\in[0,1]$ is layer-adaptive:
\[
\alpha_l \;=\; \alpha_{\min} + \gamma\,\log\!\Big(1+\frac{d_l^{\mathrm{ref}}}{d_{\mathrm{tol}}}\Big),
\quad
\]
We choose hyperparameters $\alpha_{\min}=0.01,\;\gamma=0.1,\; d_{\mathrm{tol}}=1.0$ to emphasize accuracy when degradation exceeds 1\%. 

\section{Experiment Results}
At the \textit{architecture} level, we model the proposed MRR-ONN from the chip down to the OPE by extending \textsc{Timeloop} \cite{parashar2019timeloop} and \textsc{CiMLoop} \cite{andrulis2024cimloop, andrulis2024architecture} frameworks. Photonic primitives (lasers, microring resonators, and photodiodes) are incorporated as storage and compute elements with calibrated data. All electronic and optical components are modeled based on 45nm technology and the operating frequency is set to 5 GHz. We co-optimize loop unrolling and data movement to obey optical constraints while minimizing EDP. At the \textit{algorithmic} level, we conduct behavioral simulations in \textsc{PyTorch} across alternative mapping strategies for the MRR-ONN. All networks are trained and tested using uniform 8-bit quantization of inputs, weights, and outputs.

\subsection{Towards Energy-Efficiency}

\begin{figure}[htbp]
\centerline{\includegraphics[width=\linewidth]{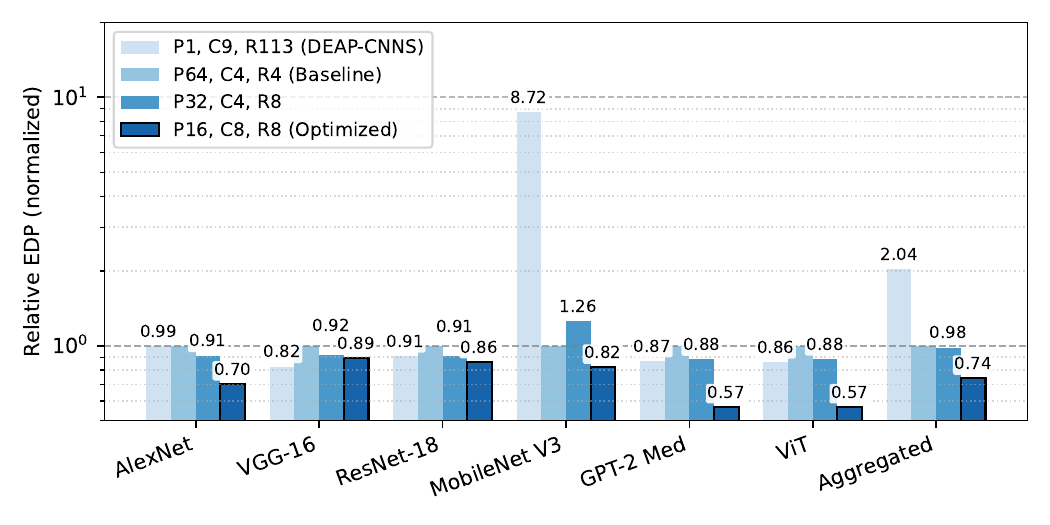}}
\caption{Energy–Delay Product (EDP) Scaling Across Neural Network Architectures}
\label{fig:edp_no_osa}
\end{figure}

\noindent\textbf{Optimized MRR-ONN Architecture Without OSA.}
We first optimize the MRR–ONN architecture \emph{without} the proposed OSA module to isolate the effects. For comparison, we include the high-channel \textsc{DEAP-CNNs} configuration $(R\!=\!113,C\!=\!9)$ and report EDP relative to a mature compact array baseline $(R\!=\!4,C\!=\!4)$ widely utilized in \cite{tait2017neuromorphic,zhao2025situ,hu2024quantized}. As shown in Fig.~\ref{fig:edp_no_osa}, the simulated relative EDP is evaluated across both convolutional models (from AlexNet to MobileNet V3) and transformer families (GPT-2 Medium and Vision Transformer). The configuration with the best aggregate rank, $(R\!=\!8,C\!=\!8)$, reduces the \emph{aggregated} relative EDP by $64\%$ versus the \textsc{DEAP-CNNs} setting and by $26\%$ versus the compact baseline, indicating that moderate-scale arrays offer the most favorable utilization and energy efficiency.

\begin{figure}[htbp]
\centerline{\includegraphics[width=\linewidth]{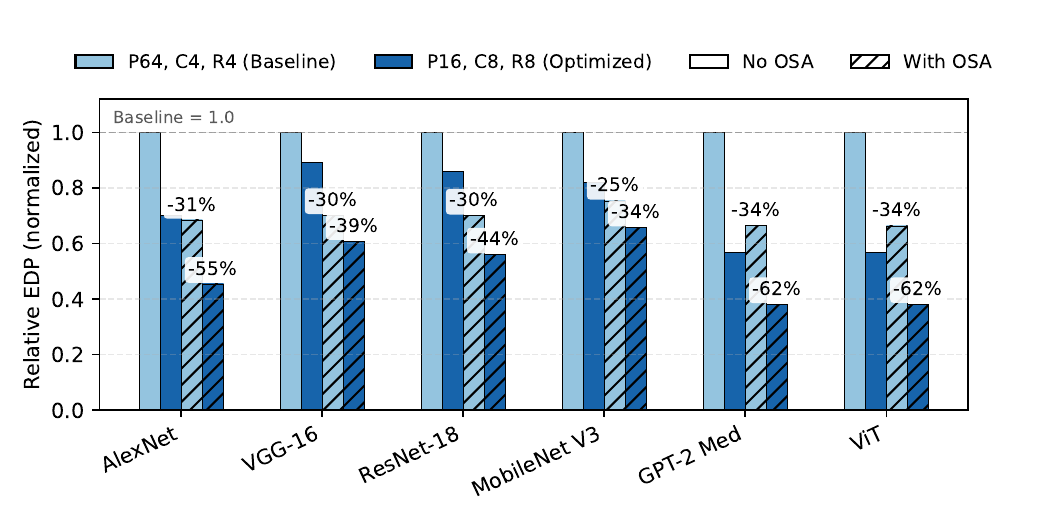}}
\caption{EDP reduction with optical shift-and-add.}
\label{fig:osa_improvement}
\end{figure}

\noindent\textbf{Impact of Optical Shift-and-Add.}
Our proposed OSA lowers EDP by restructuring multiplication as bit-serial time shifts with optical accumulation, thereby substantially reducing the energy of optical-analog and analog-digital conversion. As shown in Fig.~\ref{fig:osa_improvement}, adding OSA to the baseline MRR-ONN lowers EDP by $29\%$ relative to the no-OSA baseline, and combining OSA with optimized optical delay element (ODE) sizing yields a $37\%$ reduction over the same baseline.

\begin{figure}[htbp]
\centerline{\includegraphics[width=\linewidth]{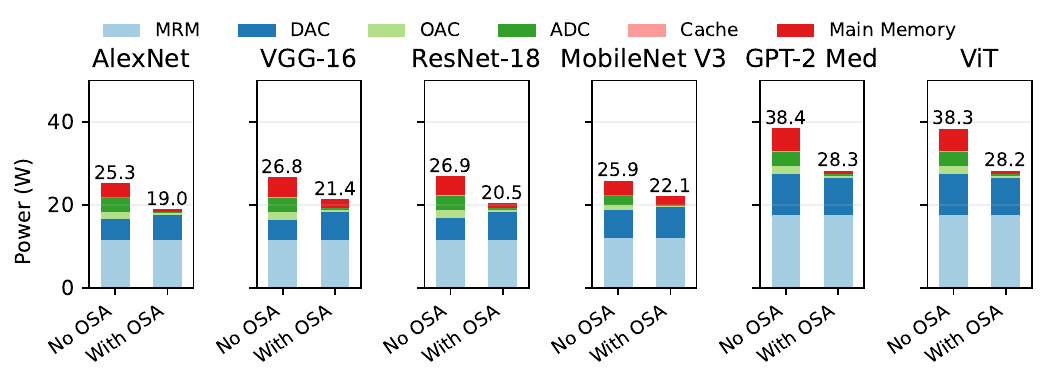}}
\caption{Power breakdown of the MRR–ONN with and without the OSA module across four workloads.}
\label{fig:osa_power}
\end{figure}

Figure~\ref{fig:osa_power} compares the component-wise power of our architecture with and without OSA. Introducing OSA reduces OAC and ADC power, and also lowers cache-memory traffic for transporting partial sums, further reducing main memory read/write energy.

\subsection{Towards Robustness: Hybrid Mapping Strategy}
\begin{figure}[htbp]
\centerline{\includegraphics[width=\linewidth]{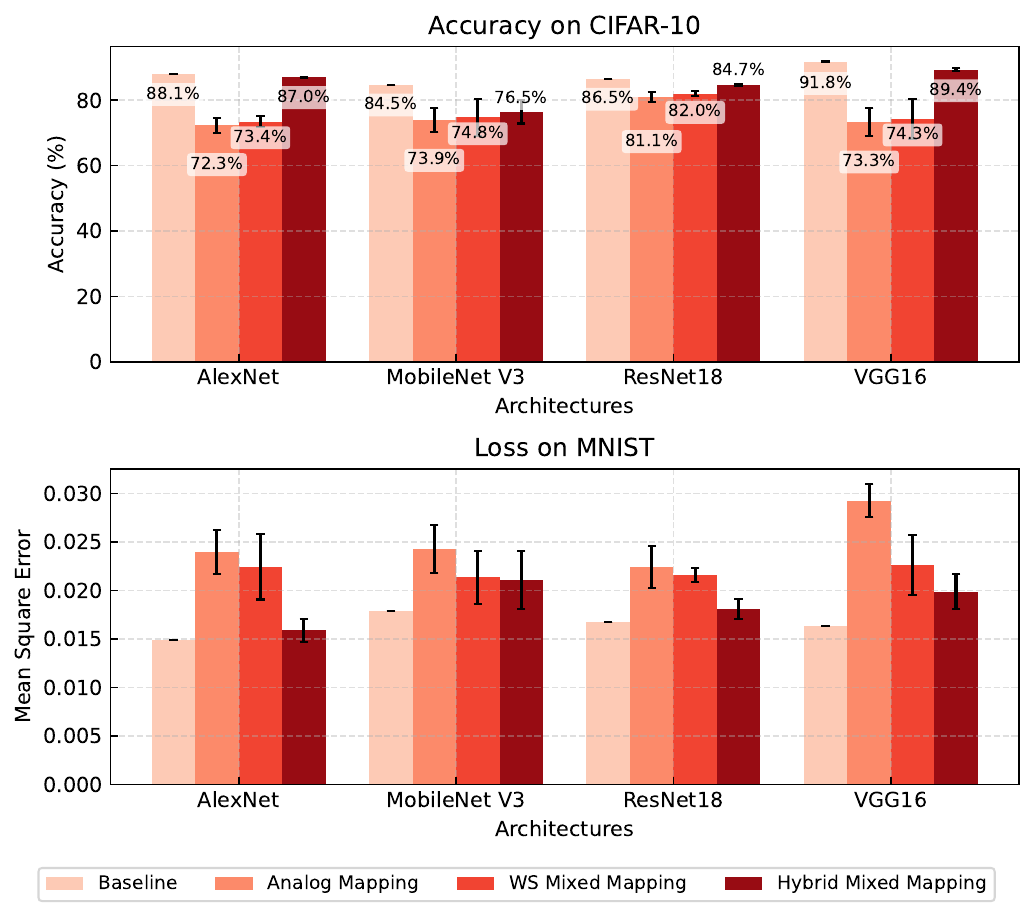}}
\caption{Accuracy and loss comparison of different mapping methods under thermal and DAC noise.}
\label{fig:hybrid-mapping-accuracy}
\end{figure}

AlexNet, MobileNet V3, ResNet-18, and VGG-16 are evaluated under DAC and thermal perturbations on \textsc{MNIST} and \textsc{CIFAR}-10 with noise standard deviations $\sigma_{\mathrm{DAC}}=0.02$ and $\sigma_{\mathrm{th}}=0.04$. The baseline is the conventional analog mapping quantized with 8-bits. The proposed digital-analog paradigm using WS mapping suppresses thermal crosstalk on time-encoded inputs. A layer-wise hybrid mapping selects between input-stationary and weight-stationary modes using the combined accuracy-EDP metric defined earlier. As shown in Fig.~\ref{fig:hybrid-mapping-accuracy} and Table~\ref{tab:EDP_accuracy_comparison}, hybrid mapping consistently outperforms WS mixed and analog mapping, yielding an average CIFAR-10 accuracy gain of $8.3\%$ and an EDP reduction of $10.8\%$ than WS. Compared to the baseline simulated accuracy without noise on PC, hybrid mapping incurs only a $3.3\%$ average accuracy loss across the four CNNs, proving that high-bit quantization levels are preserved under noise. Compared with DEAP-CNNs, the hybrid mapping with OSA and optimized ODE sizing reduces EDP by an average of $54.7\%$.

\begin{table}[t]
  \caption{Accuracy and EDP comparison between weight-stationary and hybrid mapping on CIFAR-10 dataset under thermal and DAC noise.}
  \label{tab:EDP_accuracy_comparison}
  \centering
  \small
  \begin{tabularx}{\columnwidth}{c|cc|cc}
    \toprule
    \textbf{Architectures} & \multicolumn{2}{c|}{\textbf{Accuracy} (($\%$))} & \multicolumn{2}{c}{\textbf{EDP} ($J\cdot s$)} \\  
    & \textbf{WS} & \textbf{Hybrid} & \textbf{WS} & \textbf{Hybrid} \\
    \midrule
    AlexNet & 73.4 & 87.0 ($13.6\%$↑) & 0.210 & 0.204 ($2.8\%$↓) \\
    MobileNet V3 & 74.8 & 76.5 ($2.6\%$↑)  & 0.030 & 0.026 ($13.3\%$↓) \\
    ResNet18 & 82.0 & 84.7 ($2.7\%$↑)  & 1.531 & 1.493 ($24.7\%$↓) \\ 
    VGG16 & 74.3 & 89.4 ($15.1\%$↑)  & 121.5 & 118.3 ($2.5\%$↓) \\
    \bottomrule
  \end{tabularx}
\end{table}

\section{Conclusion}
This paper demonstrates that combining optical shift-and-add with time-wavelength digital-analog MACs and noise-aware mapping significantly improves both efficiency and robustness of MRR-based ONNs. Across CNNs and Transformers, optimized arrays and OSA jointly reduce EDP while a hybrid mapping strategy mitigates DAC/thermal noise and boosts accuracy, offering a practical, scalable path toward next-generation photonic AI accelerators.



\bibliographystyle{unsrt}

\bibliography{reference}

\end{document}